\documentclass[a4paper,11pt]{article}
\pdfoutput=1 
\usepackage{jheppub} 

\usepackage[T1]{fontenc} 

\RequirePackage[displaymath]{lineno} 
\usepackage{epstopdf}
\usepackage{epsfig}
\usepackage{graphicx}
\usepackage{dcolumn}
\usepackage{bm}
\usepackage{ltablex,booktabs}
\usepackage{overpic}
\usepackage{subfigure}
\usepackage{float}
\usepackage{color}
\usepackage{amsmath}
\usepackage{mathcomp}
\usepackage{mathrsfs}
\usepackage{multirow}
\usepackage{rotating}
\usepackage{amssymb}
\usepackage{gensymb}
\usepackage{amsmath}
\usepackage{tabularx}
\usepackage{threeparttable}

\title{\boldmath Observation of $\psi(3686) \to \Xi^- K^0_S \bar{\Omega}^+ $+c.c.}

\collaborationImg{\includegraphics[width=0.15\textwidth, angle=90]{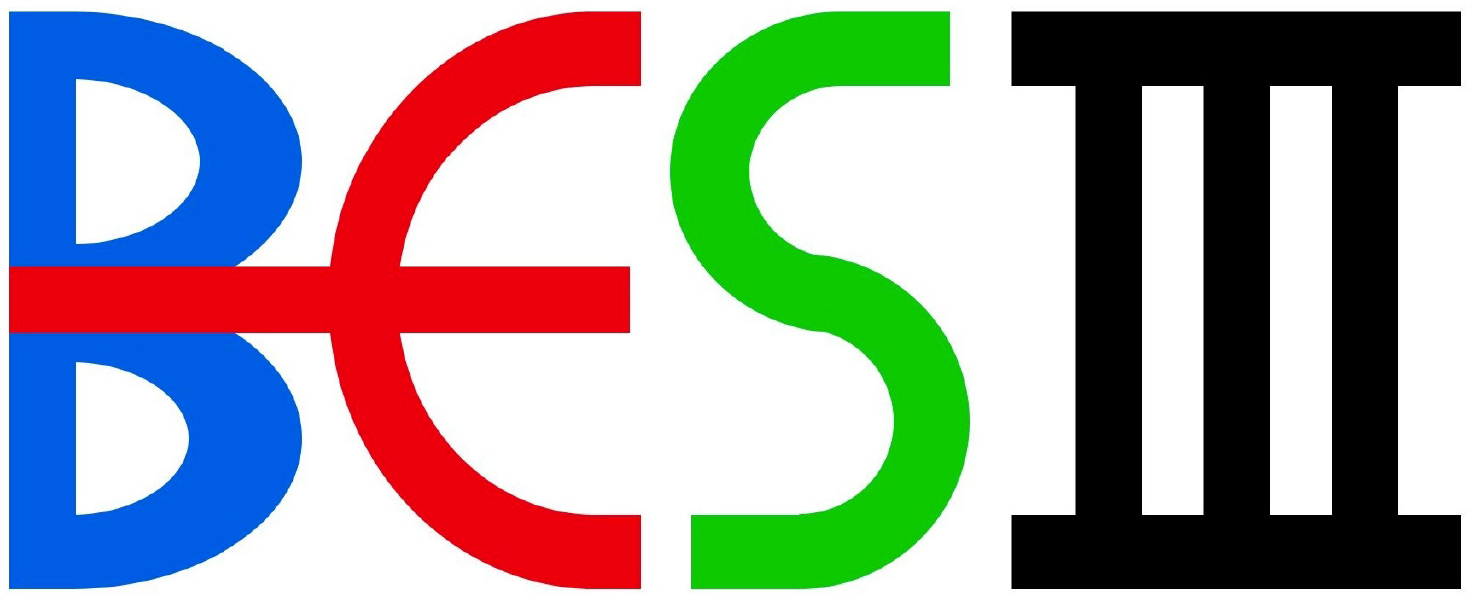}}
\collaboration{BESIII Collaboration}

\author{
M.~Ablikim$^{1}$, M.~N.~Achasov$^{4,c}$, P.~Adlarson$^{76}$, X.~C.~Ai$^{81}$, R.~Aliberti$^{35}$, A.~Amoroso$^{75A,75C}$, Q.~An$^{72,58,a}$, Y.~Bai$^{57}$, O.~Bakina$^{36}$, Y.~Ban$^{46,h}$
% ...
\\
\vspace{0.2cm}
(BESIII Collaboration)\\
\vspace{0.2cm} {\it
$^{1}$ Institute of High Energy Physics, Beijing 100049, People's Republic of China\\
% ...
$^{4}$ Budker Institute of Nuclear Physics SB RAS (BINP), Novosibirsk 630090, Russia\\
% ...
$^{35}$ Johannes Gutenberg University of Mainz, Johann-Joachim-Becher-Weg 45, D-55099 Mainz, Germany\\
$^{36}$ Joint Institute for Nuclear Research, 141980 Dubna, Moscow region, Russia\\
% ...
$^{46}$ Peking University, Beijing 100871, People's Republic of China\\
% ...
$^{57}$ Southeast University, Nanjing 211100, People's Republic of China\\
$^{58}$ State Key Laboratory of Particle Detection and Electronics, Beijing 100049, Hefei 230026, People's Republic of China\\
% ...
$^{72}$ University of Science and Technology of China, Hefei 230026, People's Republic of China\\
% ...
$^{75}$ University of Turin and INFN, (A)University of Turin, I-10125, Turin, Italy; (B)University of Eastern Piedmont, I-15121, Alessandria, Italy; (C)INFN, I-10125, Turin, Italy\\
$^{76}$ Uppsala University, Box 516, SE-75120 Uppsala, Sweden\\
% ...
$^{81}$ Zhengzhou University, Zhengzhou 450001, People's Republic of China\\

\vspace{0.2cm}
$^{a}$ Deceased\\
% ...
$^{c}$ Also at the Novosibirsk State University, Novosibirsk, 630090, Russia\\
% ...
$^{h}$ Also at State Key Laboratory of Nuclear Physics and Technology, Peking University, Beijing 100871, People's Republic of China\\
% ...

}
}

\abstract{ 
	Using a sample of $(2.712\pm0.014) \times 10^{9}$ $\psi(3686)$ events collected with the BESIII detector at the electron positron collider BEPCII, the decay $\psi(3686) \to \Xi^- K^0_S \bar{\Omega}^+ +c.c.$ is observed for the first time, which has a significance of 5.9 standard deviations.
	The branching fraction of this decay is measured to be $(2.91\pm0.47\pm0.33)\times 10^{-6}$, where the first and second uncertainties are statistical and systematic, respectively. The ratio between
	$\mathcal{B}_{\psi(3686) \to \Xi^- K^0_S \bar{\Omega}^+ +c.c.}$ and 
	$\mathcal{B}_{\psi(3686) \to \Omega^- K^+ \bar{\Xi}^0 +c.c.}$
	is determined to be 
 $1.05\pm0.23\pm0.14 $, which deviates from the isospin symmetry conservation predicted value of 0.5 by $2.1\sigma$.

}
\arxivnumber{}

\begin{document}
\maketitle
\flushbottom

\section{Introduction}
\label{sec:introduction}

The discovery of the $J/\psi$ and other charmonium(-like) states significantly impacts the development of the theory of strong interaction within the Standard Model~\cite{E598:1974sol,PhysRevLett.33.1406}. It revealed the existence of the fourth quark, known as the charm quark, while also motivating the exploration of additional heavy quarks in experimental studies. Charmonium decays serve as a critical testing ground for probing non-perturbative Quantum Chromodynamics (QCD) properties in experimental studies.~\cite{BaldiniFerroli:2019abd}.

The decay modes of the charmonium states to $B\bar{B}'P$, where $B/B^\prime$ denote a baryon and $P$ is a pseudoscalar meson, are the important modes to search for the excited baryons and provide essential information for investigating many topics involving the strong interaction, such as the color octet and singlet contributions, the violation of helicity conservation, and
 SU(3) flavor symmetry breaking effects~\cite{BaldiniFerroli:2019abd,Asner:2008nq,BESIII:2024zav}. Under the quantum number and energy conservation, all the $B\bar{B}'P$ decays of charmonium states can be summarized straightforwardly, but the branching fractions (BFs) of them are hard to predict theoretically due to the non-perturbative strong effects at low energies~\cite{Asner:2008nq}. In investigations of baryon strong interaction dynamics through charmonium decay processes, the production of at least two baryons is fundamentally required by baryon number conservation. To probe low-energy baryon excited states, the inclusion of an additional pseudoscalar meson becomes experimentally advantageous due to its relatively low mass. Furthermore, three-body decays typically exhibit larger BFs and simpler dynamics compared to higher-multi-body decays (four-body or five-body decays). Therefore, the studies of $B\bar{B}'P$ decays of charmonium states in experiments have become a necessary task.

Recently, the first observation of the decay $\psi(3686) \to \Xi^0 K^- \bar{\Omega}^+ +c.c.$ has been reported and the corresponding BF has been measured by the BESIII collaboration~\cite{BESIII:2024zav}.
This study expanded our knowledge of the decay mechanism of $\psi(3686)$ and provided an ideal environment to search for possible $\Xi^*$ and $\Omega^*$ states~\cite{BESIII:2020lkm,BESIII:2023olq}. The research of its isospin partner channels can also search for the possible baryon excited states and test the SU(3) flavor symmetry, which is interesting.

In this paper, the first observation of the decay $\psi(3686) \to \Xi^- K^0_S \bar{\Omega}^+ +c.c.$ is reported and the corresponding
BF is measured using $(2.712 \pm 0.014) \times 10^{9}$
$\psi(3686)$ events~\cite{BESIII:2017tvm} collected with the BESIII detector.
In addition, the possible baryon excited states
 are searched for and the conservation of isospin symmetry is tested in this decay. Throughout this paper, the charge-conjugate mode is always implied. 

\section{BESIII detector and Monte Carlo simulation}
\label{sec:BESIII and MC}
The BESIII detector~\cite{BESIII:2009fln} records symmetric $e^+e^-$ collisions 
provided by the BEPCII storage ring~\cite{Yu:2016cof}
in the center-of-mass energy range from 1.84 to 4.95~GeV,
with a peak luminosity of $1.1 \times 10^{33}\;\text{cm}^{-2}\text{s}^{-1}$ 
achieved at $\sqrt{s} = 3.773\;\text{GeV}$. 
BESIII has collected large data samples in this energy region~\cite{BESIII:2020nme, Zhang:2022bdc}. The cylindrical core of the BESIII detector covers 93\% of the full solid angle and consists of a helium-based
 multilayer drift chamber~(MDC), a time-of-flight
system~(TOF), and a CsI(Tl) electromagnetic calorimeter~(EMC),
which are all enclosed in a superconducting solenoidal magnet
providing a 1.0~T magnetic field.
The solenoid is supported by an
octagonal flux-return yoke with resistive plate counter muon
identification modules interleaved with steel. The charged-particle momentum resolution at $1~{\rm GeV}/c$ is $0.5\%$, and the ${\rm d}E/{\rm d}x$ resolution is $6\%$ for electrons
from Bhabha scattering. The EMC measures photon energies with a resolution of $2.5\%$ ($5\%$) at $1$~GeV in the barrel (end cap) region. The time resolution in the plastic scintillator TOF barrel region is 68~ps, while that in the end cap region was 110~ps. The end cap TOF system was upgraded in 2015 using multigap resistive plate chamber technology, providing a time resolution of 60~ps, which benefits 83\% of the data used in this analysis~\cite{Li:2017jpg, Guo:2017sjt,  Cao:2020ibk}.

Monte Carlo (MC) simulated data samples produced with a {\sc geant4}~\cite{GEANT4:2002zbu} based software package, which includes the geometric description of the BESIII detector and the detector response, are used to optimize the event selection criteria, estimate the signal efficiency and background level.The simulations incorporate the beam-energy spread and initial-state radiation in the $e^+e^-$ annihilation using the generator {\sc kkmc}~\cite{Jadach:2000ir}. The inclusive MC sample includes the production of the $\psi(3686)$ resonance, the initial-state radiation production of the $J/\psi$ meson, and the continuum processes incorporated in {\sc kkmc}~\cite{Jadach:2000ir}. Particle decays are generated by {\sc evtgen}~\cite{Lange:2001uf, Ping:2008zz} for the known decay modes with BFs taken from the Particle Data Group~\cite{PDG} and {\sc lundcharm}~\cite{Chen:2000tv, Yang:2014vra} for the remaining unknown ones. Final-state radiation from charged final-state particles is included using the {\sc photos} package~\cite{Richter-Was:1992hxq}. The inclusive MC sample at the $\psi(3686)$ resonance, consisting of $2.712 \times 10^9$ events, is analysed with a generic event-type examination tool, TopoAna~\cite{Zhou:2020ksj}, to identify potential backgrounds. To determine the detection efficiency, a signal MC sample comprising 2 million events of the signal decay chain of $\psi(3686) \to \Xi^- K^0_S \bar{\Omega}^+, \Xi^- \to \Lambda(\to p \pi^-) \pi^-, K^0_S \to \pi^+\pi^-$ is generated uniformly distributed in phase space, along with inclusive $\bar{\Omega}^+$ decays. The data sample collected at the center-of-mass energies of 3.650 and 3.773 GeV, corresponding to total integrated luminosities of 410~pb$^{-1}$ and 7.93~fb$^{-1}$~\cite{Ablikim:2013ntc}, are used to estimate the continuum production contribution. The data sample of $(2.712\pm0.014) \times 10^{9}$ $\psi(3686)$ events is used to study  $\psi(3686) \to \Xi^- K^0_S \bar{\Omega}^+$.

\section{Event selection}
\label{sec:event selection}
As the full reconstruction method suffers from low detection efficiency, a partial-reconstruction strategy is applied, in which only the $\Xi^-$ and $K^0_S$ candidates are reconstructed, without identifying $\bar{\Omega}^+$. The cascade decay of interest is $\psi(3686) \to \Xi^- K^0_S \bar{\Omega}^+$,~with $\Xi^- \to \Lambda \pi^-$, $\Lambda \to p \pi^- $ and $K^0_S \to \pi^+\pi^-$. 
 Charged tracks detected in the MDC are required to be within a polar angle ($\theta$) range of $|\rm{cos\theta}|<0.93$, where $\theta$ is defined with respect to the $z$-axis, which is the symmetry axis of the MDC. For these tracks, the distance of closest approach to the interaction point (IP) is required to be less than 20~cm along the MDC axis.
Particle identification~(PID) for charged tracks combines measurements of the energy deposited in the MDC~(d$E$/d$x$) and the flight time in the TOF to form likelihoods $\mathcal{L}(h)~(h=p,K,\pi)$ for each hadron $h$ hypothesis.
Tracks are identified as protons when the proton hypothesis has the greatest likelihood ($\mathcal{L}(p)>\mathcal{L}(K)$ and $\mathcal{L}(p)>\mathcal{L}(\pi)$), while charged pions are identified by comparing the likelihoods for the kaon and pion hypotheses, $\mathcal{L}(\pi)>\mathcal{L}(K)$.
PID is performed for the proton from $\Lambda$ and the pion from $\Xi^-$. The pion from $\Lambda$ can be effectively identified through the vertex fit of $\Lambda$, whereas the selection of pion from $\Xi^-$ through vertex fit of $\Xi^-$ is not satisfactory. PID applied on the pion from $\Xi^-$ can suppress the background of misidentification of it.

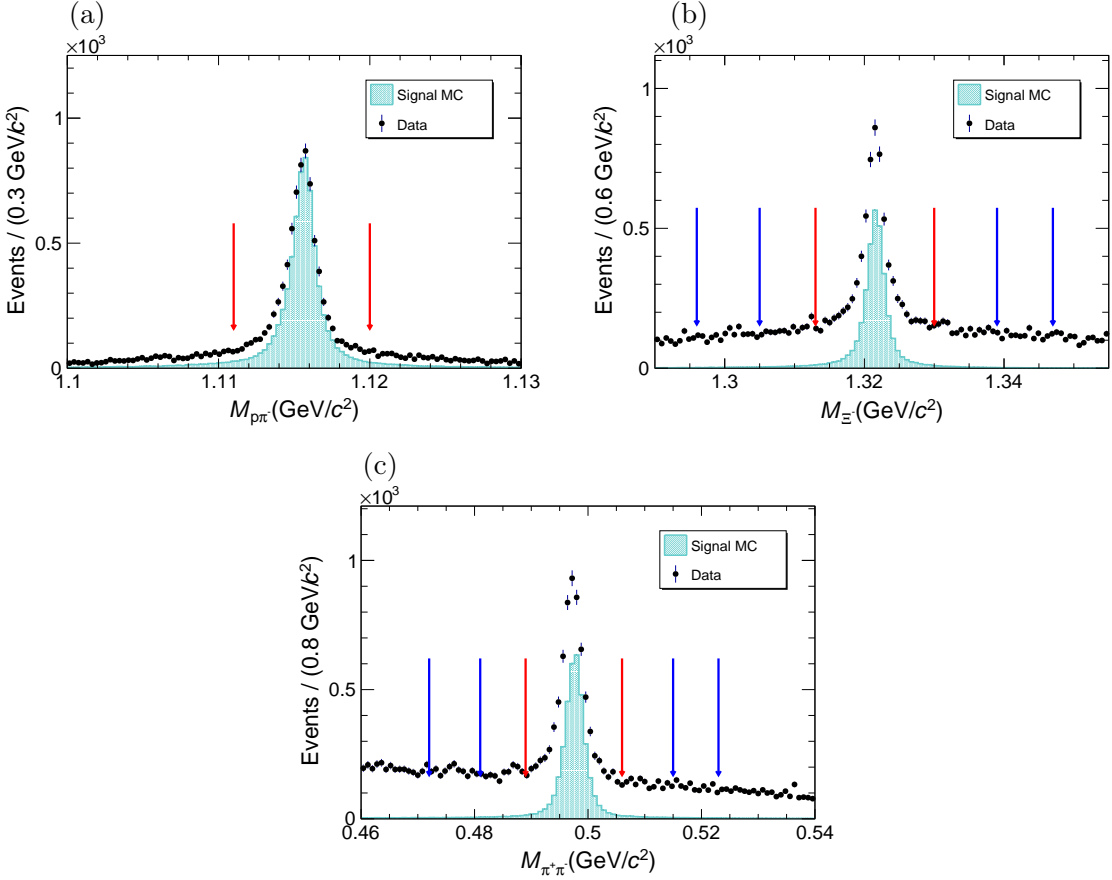
\begin{figure}[htbp]
	\begin{center}
        \mbox{
            \put(-230, 10){
                \begin{overpic}[width = 0.5\linewidth]{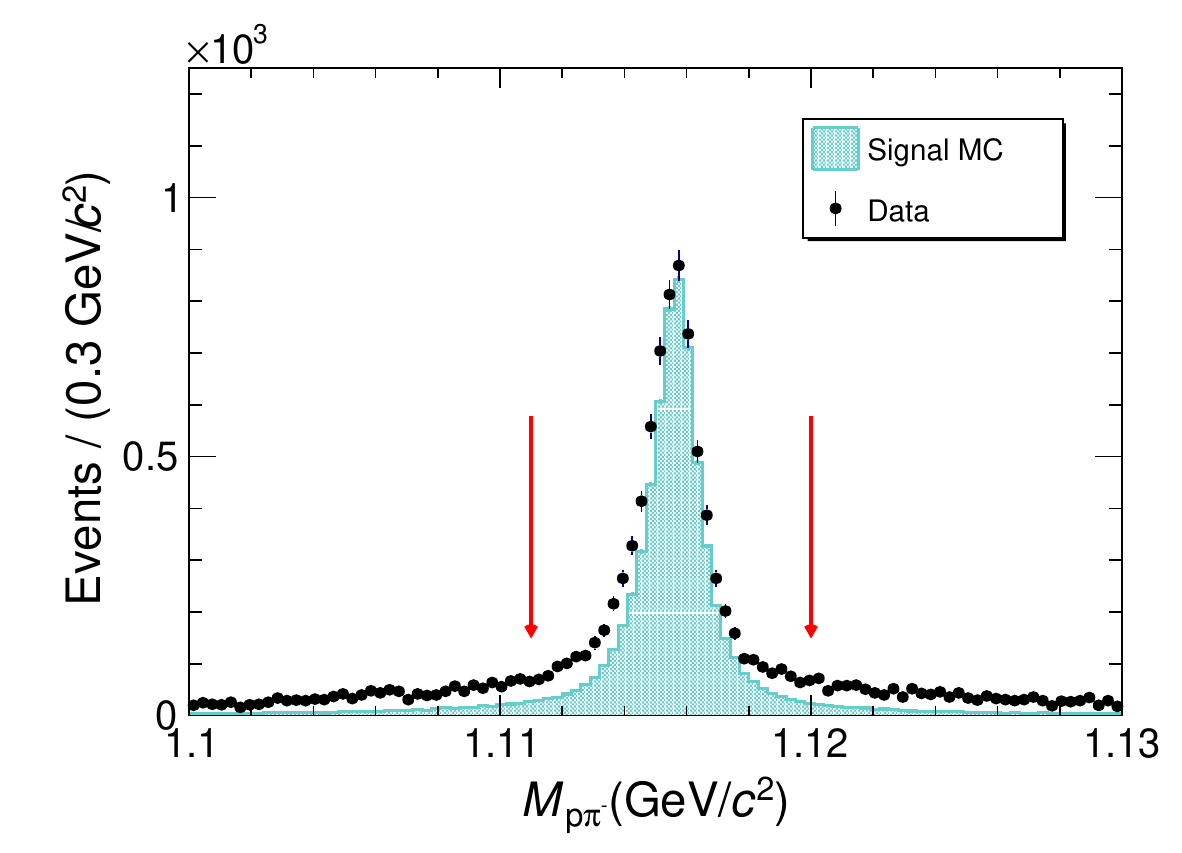}
                \end{overpic}}
               
            \put(-10, 10){
                \begin{overpic}[width = 0.5\linewidth]{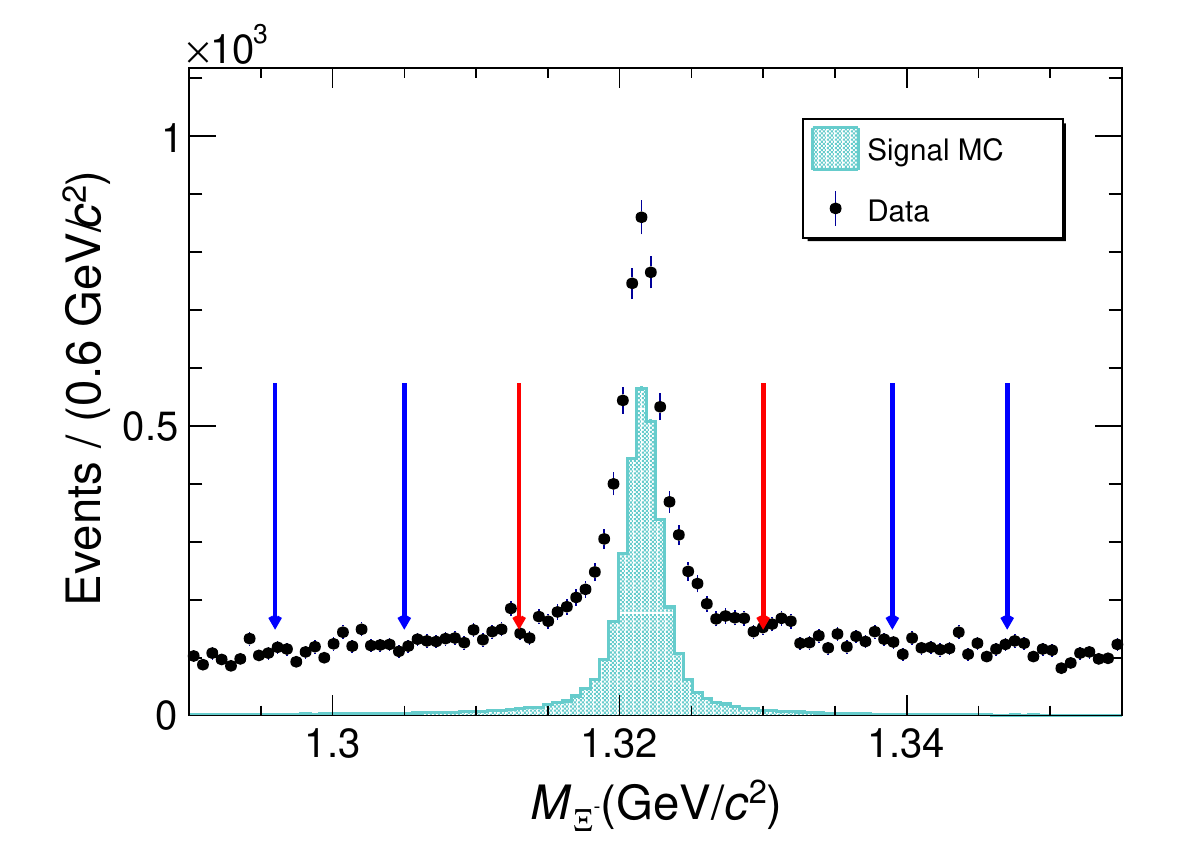}
                \end{overpic}}
             \put(-120, -160){
            	\begin{overpic}[width = 0.5\linewidth]{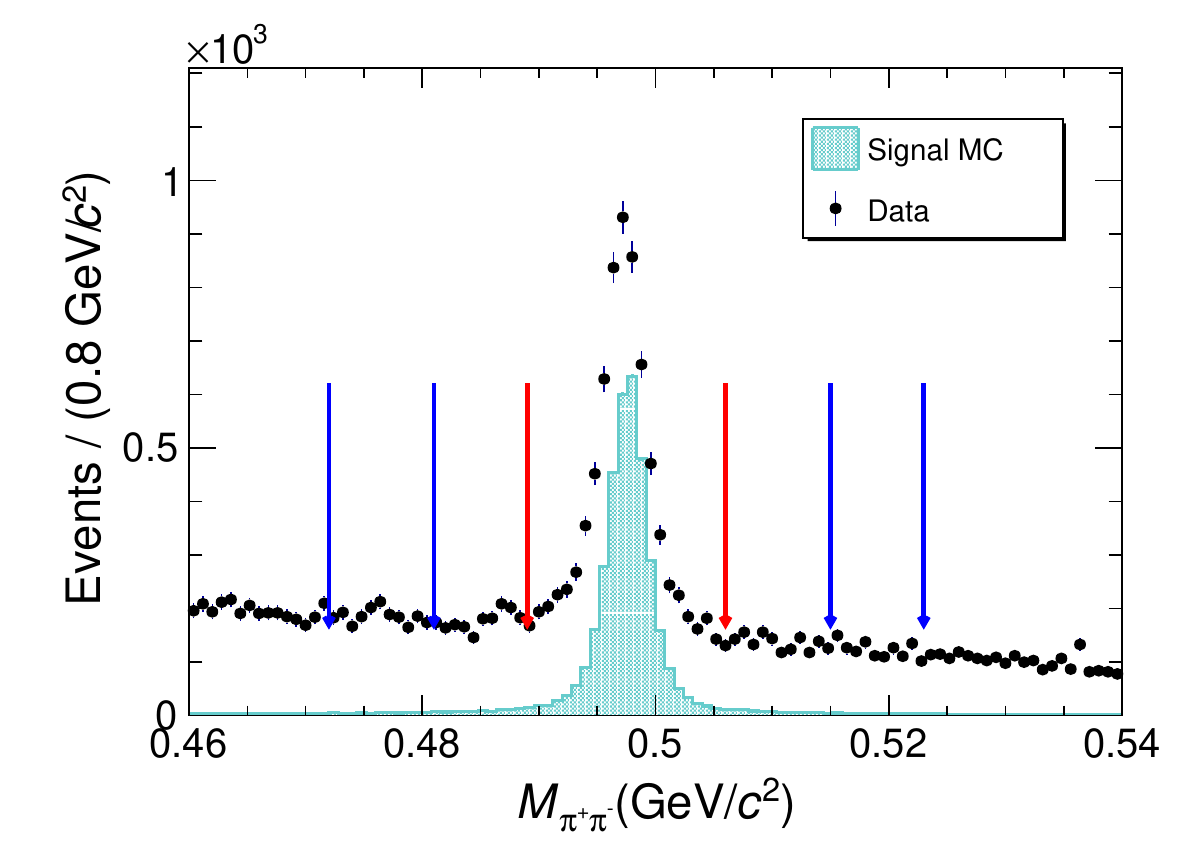}
            \end{overpic}}
            \put(-195, 165) { (a)}
            \put(30, 165)   { (b)}
             \put(-85, -5)   { (c)}
        }
	\end{center}
	\caption{
The distributions of $M_{p \pi^-}$(a), $M_{\Xi^-}$(b) and $M_{\pi^+\pi^-}$(c) for the data and signal MC sample. The red arrows show the signal region, and the blue arrows show the sideband regions. 
	}
	\label{Lambda and Omega}
\end{figure}

The $\Lambda$ candidates are reconstructed using vertex fits~\cite{Xu:2009zzg} on $p \pi^-$ pairs with the requirement $\chi^{2}<200$. The $p \pi^-$ invariant mass ($M_{p \pi^-}$) must be within the $\Lambda$ signal region, $M_{p \pi^-} \in [1.111, 1.120]$~GeV/$c^2$, as shown in Fig.~\ref{Lambda and Omega}(a). The signal region corresponds to six times the $\Lambda$ mass resolution, as determined by fitting the distribution of $M_{p\pi^-}$ for the signal MC sample.

The $\Xi^-$ candidate is reconstructed with a $\Lambda$ candidate and a $\pi^-$ by another vertex fit. 
The $d_{\Xi^-}$, which is the decay length of the $\Xi^-$ obtained by the vertex fit, is required to be larger than 0. If there is more than one $\Xi^-$ candidate, the one with the minimum $\sqrt{(\frac{M_{\Lambda \pi^-}-M_{\Xi^-}^{\rm PDG}}{\sigma_{M_{\Lambda\pi^-}}})^2+(\frac{M_{p \pi^-}-M_{\Lambda}^{\rm PDG}}{\sigma_{M_{p\pi^-}}})^2}$ is chosen, where $M_{\Xi^-}^{\rm PDG}$ and $M_{\Lambda}^{\rm PDG}$ are the nominal masses of $\Xi^-$ and $\Lambda$ cited from the Particle Data Group~\cite{PDG}. $\sigma_{M_{\Lambda\pi^-}}$ and $\sigma_{M_{p\pi^-}}$ are the mass resolution of $\Xi^-$ and $\Lambda$, which is 2.8 and 1.5 MeV/$c^2$ respectively.
The invariant mass of the $\Xi^-$ candidate is defined as $M_{\Xi^-}=M_{\Lambda \pi^-}-M_{p \pi^-}+M_{\Lambda}^{\rm PDG}$, which is used to improve the mass resolution of  the $\Xi^-$ candidate. The distribution of $M_{\Xi^-}$ is shown in Fig.~\ref{Lambda and Omega}(b). The $\Xi^-$ signal region is defined as $M_{\Xi^-}$ $\in$ $[1.313, 1.330]$~GeV/$c^2$, corresponding to six times the $\Xi^-$ mass resolution determined by fitting the distribution of $M_{\Xi^-}$ for the signal MC sample. The sideband regions defined as $M_{\Xi^-} \in ([1.296, 1.305]$ $\cup$ $[1.339, 1.347])$~GeV/$c^2$ are used to investigate the background.

Each $K_{S}^0$ candidate is reconstructed from two oppositely charged tracks satisfying $|V_{z}|<$ 20~cm.
The two charged tracks are assigned
as $\pi^+\pi^-$ without imposing further PID criteria. They are constrained to
originate from a common vertex. The
decay length of the $K^0_S$ candidate is required to be greater than
twice the vertex resolution away from the IP.
The quality of the vertex fits is ensured by a requirement on the $\chi^2_{st}$ and $\chi^2_{nd}$ ($\chi^2_{st}<200$ and $\chi^2_{nd}<200$), which $\chi^2_{st}$ and $\chi^2_{nd}$ are the Chi-Square of primary and secondary vertex fit, respectively. The distribution of $M_{\pi^+\pi^-}$ is shown in Fig.~\ref{Lambda and Omega}(c). The $K^0_S$ signal region is defined as $M_{\pi^+\pi^-}$ $\in$ $[0.489, 0.506]$~GeV/$c^2$, corresponding to six times the $K^0_S$ mass resolution determined by fitting the distribution of $M_{\pi^+\pi^-}$ for the signal MC sample. The sideband regions defined as $M_{\pi^+\pi^-} \in([0.472, 0.481]$ $\cup$ $[0.515, 0.523])$~GeV/$c^2$ are used to study the background.

Signal events manifest themselves through an $\bar{\Omega}^+$ peak in the distribution of the invariant mass recoiling against the $\Xi^- K^0_S$ system($RM_{\Xi^- K^0_S}$), $RM_{\Xi^- K^0_S} = \sqrt{|p_{e^+ e^-} - p_{\Xi^-} -p_{K^0_S}|^2}$, where $p_{e^+ e^-}$, $p_{\Xi^-}$ and $p_{K^0_S}$ refer to the 4-momentum of the initial $e^+ e^-$, $\Xi^-$ and $K^0_S$.

\section{Detection efficiency}
\label{sec:detection efficiency}

\begin{figure*}[htbp]
    \begin{center}
        \mbox{
            \put(-230, 10){
                \begin{overpic}[width = 0.5\linewidth]{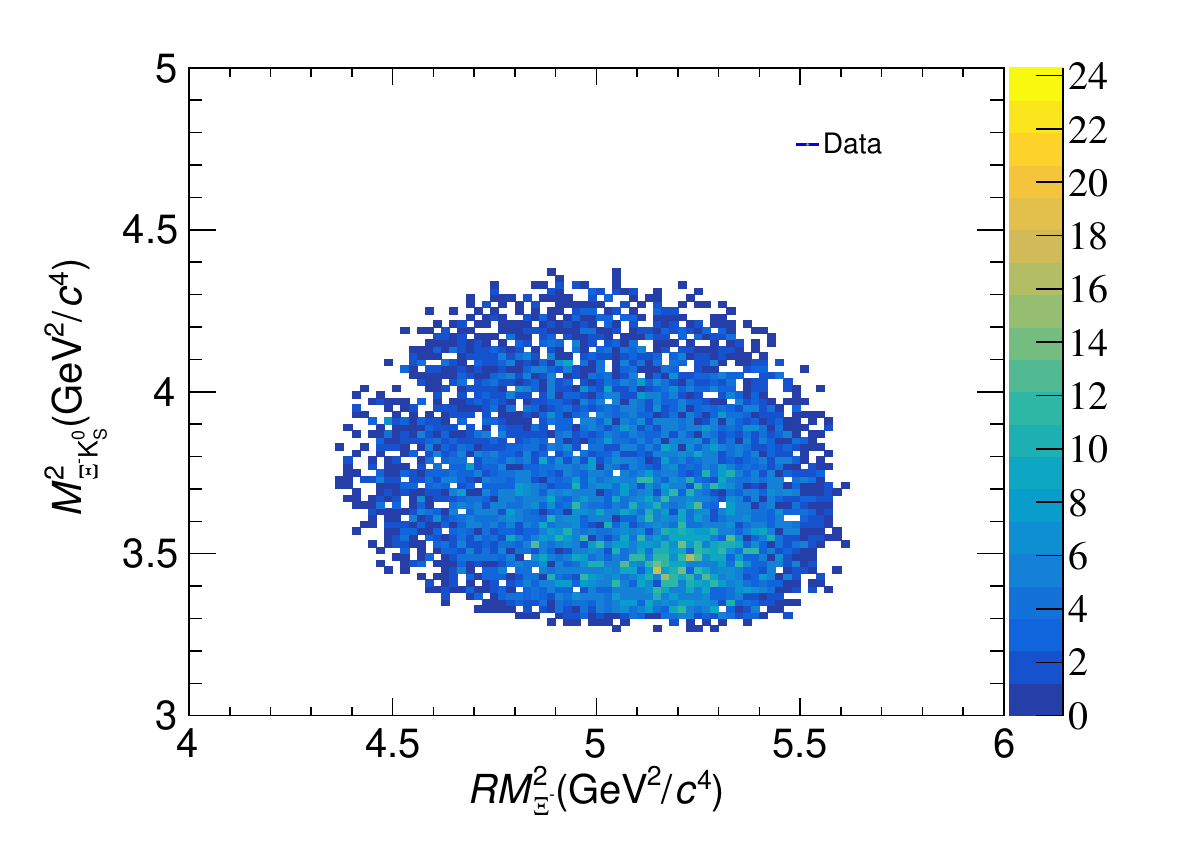}
                \end{overpic}}
            \put(-10, 10){
                \begin{overpic}[width = 0.5\linewidth]{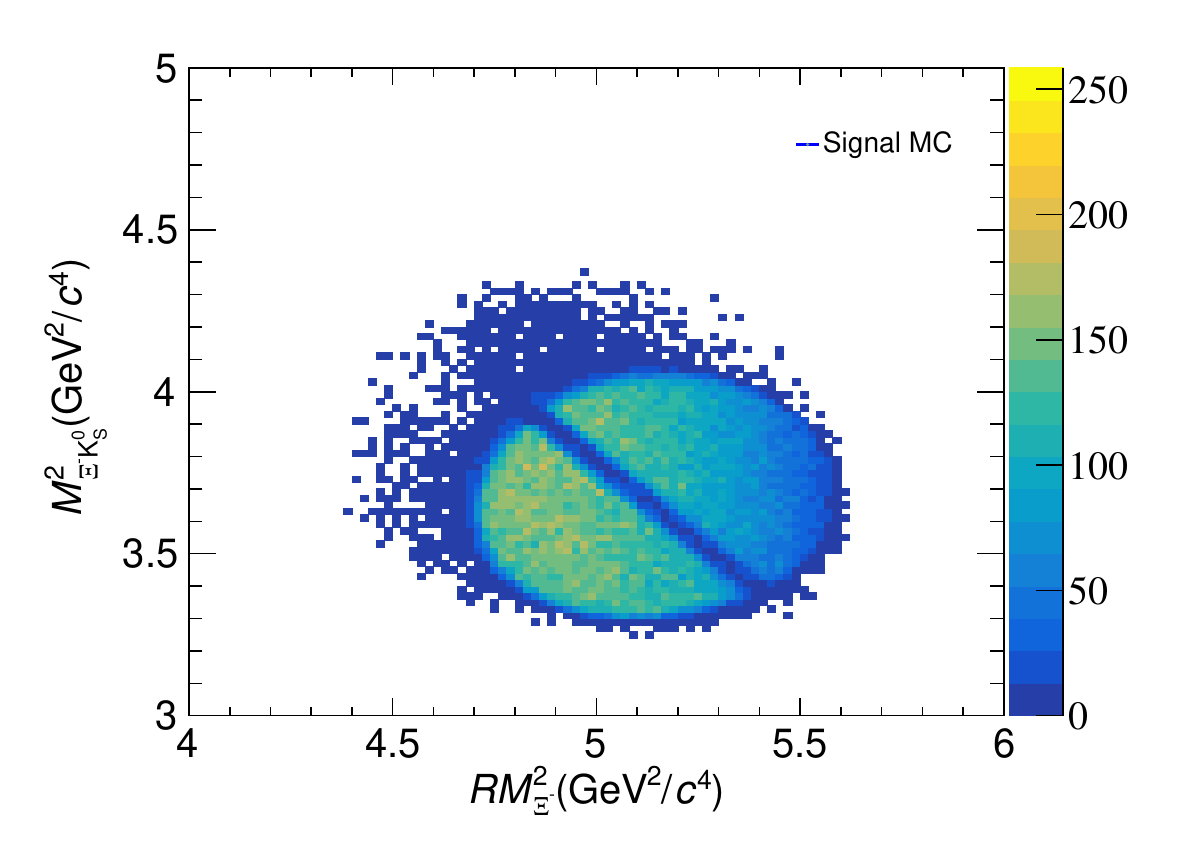}
                \end{overpic}}
        
            \put(-195, 160) { (a)}
            \put(30, 160)   { (b)}

        }
    \end{center}
    \caption{The Dalitz plot in the signal regions for the data~(a) and signal MC sample~(b).
    }
    \label{fig:Dalitz Plot}
\end{figure*}

The detection efficiency is determined with signal MC simulation. Thus, it is necessary to assess the potential impact of intermediate states on the efficiency. As shown in Fig.~\ref{fig:Dalitz Plot} exemplarily, which provides the Dalitz plot for the signal regions, no intermediate state, $\Omega^{*-}$~($\Xi^-K^0_S$) or $\bar{\Xi}^{*+}$~($\bar{\Omega}^+K^0_S$), is evident in the data sample. Therefore, the efficiency determined by the signal MC simulation is acceptable. The diagonal band observed in the Dalitz plot of the signal MC arises from the requirement of the recoil mass of $K^0_S$($RM_{\pi^+\pi^-}$) to veto the process $\psi(3686)\to\pi^+\pi^-J/\psi$. $RM_{\pi^+\pi^-}$ is defined as $RM_{\pi^+\pi^-} = \sqrt{|p_{e^+ e^-} - p_{K^0_S}|^2}$. The specific requirement will be mentioned in the Sec.~\ref{sec:background study}. In contrast, this band does not appear in the Dalitz plot of the data due to the influence of background effects.

\section{Background study}
\label{sec:background study}
According to the study of the inclusive MC sample, there are peaking $J/\psi$ background events in the recoiling mass distribution against $K^0_S$($RM_{\pi^+\pi^-}$), which is from the $\psi(3686) \to \pi^+\pi^- J/\psi$ process. We require $RM_{\pi^+\pi^-}$ to be less than 3.09~GeV/$c^2$ or larger than 3.105~GeV/$c^2$ to veto such background.

\begin{figure*}[htbp]
    \begin{center}
        \mbox{
            \put(-230, 10){
                \begin{overpic}[width = 0.5\linewidth]{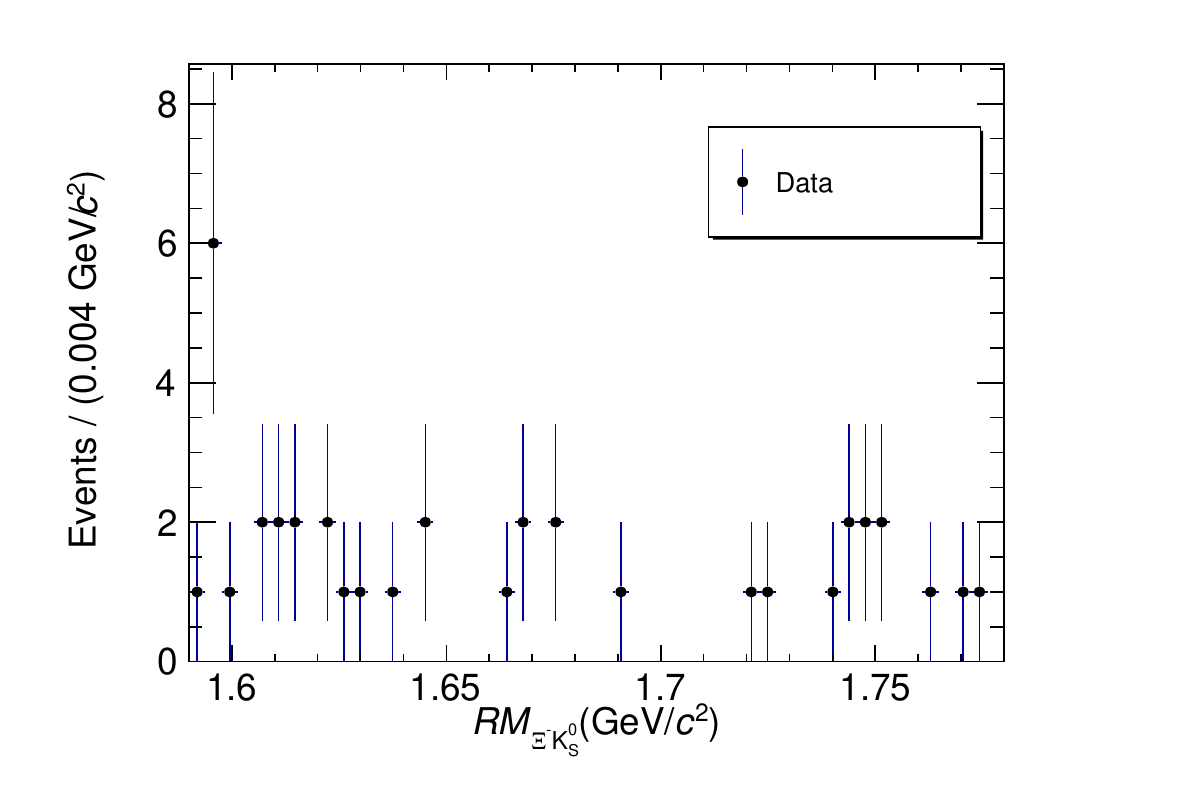}
                \end{overpic}}
            \put(-10, 10){
                \begin{overpic}[width = 0.5\linewidth]{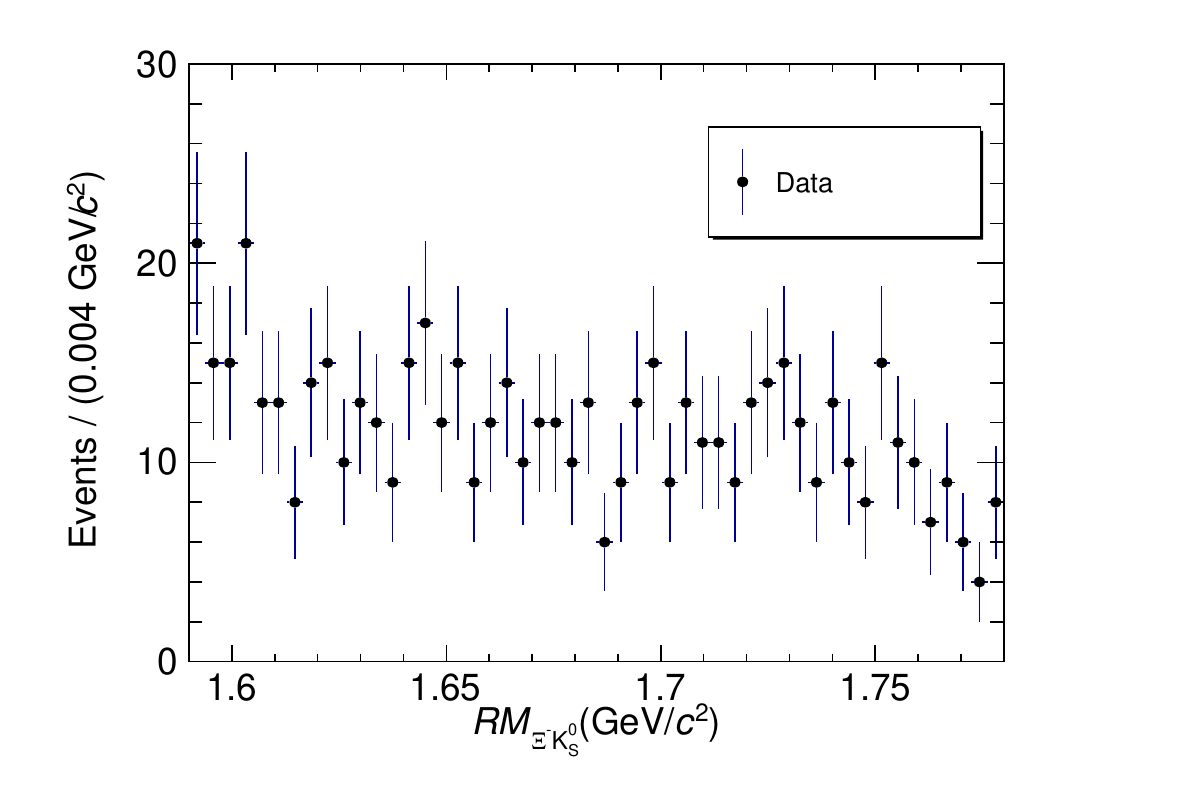}
                \end{overpic}}
        
            \put(-195, 150) { (a)}
            \put(30, 150)   { (b)}

        }
    \end{center}
    \caption{The distributions of $RM_{\Xi^-K^0_S}$ for the data at the center-of-mass energies $\sqrt{s}=3.650$ GeV~(a) and $\sqrt{s}=3.773$ GeV~(b).
    }
    \label{fig:cotinue Plot}
\end{figure*}

Further studies are performed on the surviving events in the $\bar{\Omega}^+$ signal region from the inclusive MC sample, on the events in the $\Xi^-$ mass sideband regions from data, and on the events in the $K^{0}_{S}$ sideband regions from data. These investigations indicate that there is no significant source of peaking background in the $RM_{\Xi^-K^{0}_{S}}$ distribution. To investigate the contamination from continuum processes~\cite{BESIII:2017tvm}, the same selection criteria are applied to the data samples at the center-of-mass energies $\sqrt{s}=3.650$ GeV and $\sqrt{s}=3.773$ GeV. Few events from these sample survived and do not contribute a peaking structure, which are shown in Fig.~\ref{fig:cotinue Plot}, indicating the continuum background neglected.

\section{Signal yield and BF}
\label{sec:BF}
\begin{figure*}[htbp]
    \begin{center}
        \mbox{
            \put(-120, 10)
            {
                \begin{overpic}[width = 0.5\linewidth]{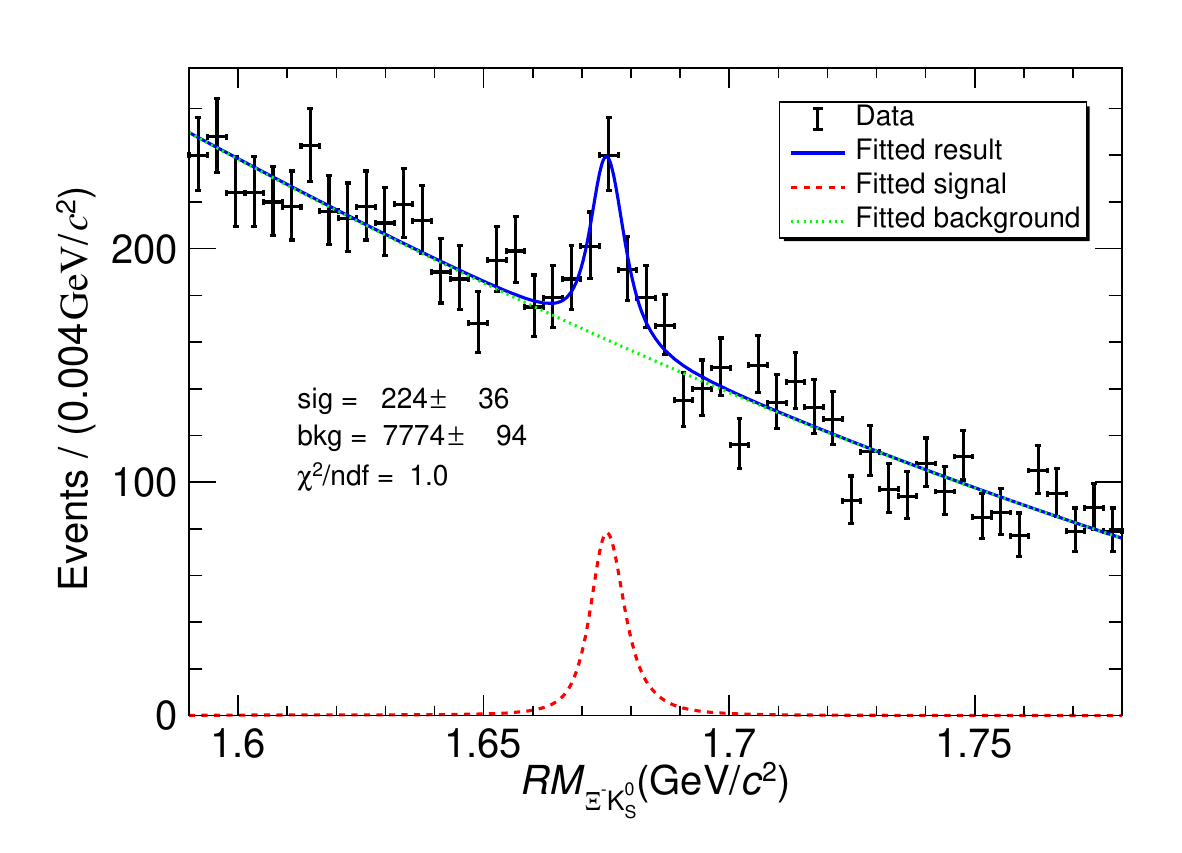}
                \end{overpic}
            }
        }
    \end{center}
    \caption{
     Fit to the $RM_{\Xi^- K^0_S}$ distribution of the accepted candidates in $\psi(3686)$ data.
}
    \label{fig:RMOmegaK_fit}
\end{figure*}

To determine the signal yield, an unbinned maximum-likelihood fit is performed on the $RM_{\Xi^- K^{0}_{S}}$ distribution~\cite{BESIII:2018gvg}. In the fit, the signal shape is described by the signal MC shape convolved with a Gaussian function with free parameters mean value and $\sigma$, where the Gaussian function is used to account for the difference in mass resolution between data and MC simulation. The signal MC shape is the $RM_{\Xi^- K^{0}_{S}}$ shape extracted from the signal MC sample by using the RooKeysPdf in RooFit toolkit~\cite{Verkerke:2003ir} The background shape is described by a second-order Chebyshev polynomial function. The fit result is shown in Fig.~\ref{fig:RMOmegaK_fit}. 
The signal yield from the fit is $N_\textrm{obs.} = 224 \pm 36$.
The statistical significance of the $\Bar{\Omega}^+$ signal is $6.1\sigma$, which is determined from the change in the log-likelihood values and the corresponding change in the number of degrees of freedom with and without including the signal contribution in the fit.

The BF of the $\psi(3686) \to \Xi^- K^0_S \bar{\Omega}^+$ decay is calculated as
\begin{footnotesize}
    \begin{equation}
        \begin{aligned}
         \mathcal{B}_{\psi(3686) \to \Xi^- K^0_S \bar{\Omega}^+ } = \frac{N_\textrm{obs.}} {N_{\psi(3686)} \cdot \mathcal{B}_{\Xi^- \to \Lambda \pi^-} \cdot \mathcal{B}_{\Lambda \to p \pi^-} \cdot \mathcal{B}_{K^0_S \to \pi^+ \pi^-} \cdot \epsilon},
         \end{aligned}
    \end{equation}
\end{footnotesize}
\setlength{\parindent}{0pt}

where $N_{\psi(3686)}$ is the total number of $\psi(3686)$ events~\cite{BESIII:2017tvm}, and $\epsilon=6.43\%$ is the detection efficiency. The efficiency uncertainty is considered as a systematic uncertainty, which is estimated in the Sec.~\ref{sec:systematic uncertainty}. $\mathcal{B}_{\Xi^- \to \Lambda \pi^-}$,  $\mathcal{B}_{\Lambda \to p \pi^-}$ and $\mathcal{B}_{K^0_S \to \pi^+ \pi^-}$ are the BFs of $\Xi^- \to \Lambda \pi^-$, $\Lambda \to p \pi^-$ and $K^0_S \to \pi^+ \pi^-$ decays, respectively, cited from the PDG~\cite{PDG}. With these inputs, the BF of $\psi(3686) \to \Xi^- K^{0}_{S} \bar{\Omega}^+ $ is determined to be $(2.91\pm0.47)\times10^{-6}$, where the uncertainties is statistical.

\section{Systematic uncertainty}
\label{sec:systematic uncertainty}
The systematic uncertainties in the  $\mathcal{B}_{\psi(3686) \to \Xi^- K^0_S \bar{\Omega}^+}$ measurement include contributions associated with the pion-tracking, PID, $\Lambda$ reconstruction, the requirement on $M_{\Xi^-}$ and $d_{\Xi^-}$, $\psi(3686) \to \pi^+\pi^- J/\psi$ veto, $K^0_S$ reconstruction, signal and background shapes, fit bias, MC generator, the size of signal MC sample, the input BFs~\cite{PDG}, and the total number of $\psi(3686)$ events~\cite{BESIII:2017tvm}.

\hspace{0.7cm}The systematic uncertainties arising from the pion-tracking and PID are studied with the well understood decays $J/\psi \to p \overline{p} \pi^+ \pi^-$~\cite{Yuan:2015wga} and $\psi'\to \gamma \chi_{cJ}, \chi_{cJ}\to \gamma \rho^0(\omega)$~\cite{BESIII:2011ysp}, and both are assigned as 1.0\% per track. 

\hspace{0.7cm}The systematic uncertainty associated with the $\Lambda$-reconstruction includes effects from the tracking and PID for the proton and the tracking for pion, and the requirement on $M_{p\pi^-}$. 
This uncertainty is estimated with a control sample of $J/\psi \to p K^- \bar{\Lambda}$ decays~\cite{BESIII:2023drj,BESIII:2024nif}.
The momentum-dependent ratios of the $\Lambda$ reconstruction efficiencies between data and MC simulation are used to re-weight the signal MC sample. The difference between the baseline and reweighted detection efficiencies, 2.7\%, is taken as the systematic uncertainty.

\hspace{0.7cm}The systematic uncertainties associated with the requirements on $M_{\Xi^-}$ and $d_{\Xi^-}$ are studied with a control sample of $J/\psi \to \Xi^- \bar{\Xi}^{+}$, where $\Xi^-$ is fully reconstructed with $\Xi^-\to \Lambda \pi^-$, $\Lambda \to p\pi^-$. The reconstruction strategy for $\Xi^-$ is the same as in section~\ref{sec:event selection} and the uncertainty is determined to be 0.2\%.

\hspace{0.7cm}The $\psi(3686) \to \pi^+\pi^- J/\psi$ background is vetoed by requiring $RM_{\pi^+\pi^-}$ to be less than 3.09 GeV/$c^2$ or more than 3.105 GeV/$c^2$. By changing the requirement of $RM_{\pi^+\pi^-}$ to be less than 3.08 GeV/$c^2$ or more than 3.115 GeV/$c^2$, the change of the re-measured BF, 2.4$\%$, is assigned as the systematic uncertainty.

 \hspace{0.7cm}Two control samples $J/\psi \to K^*(892)^-K^+,K^*(892)^-\to K^0_S \pi^-$ and $J/\psi \to \phi K^0_S K^+\pi^-$ are used to study the systematic uncertainty of $K^0_S$ reconstruction. By comparing the inconsistency in the data and signal MC regarding the reconstruction of $K^0_S$, this uncertainty is estimated to 1.2\%.

\hspace{0.7cm}The pseudo-experiment method is used to estimate the systematic uncertainty related to the fit bias, signal shape and background shape. 
The probability density function (PDF) is utilized to describe the signal and background distributions derived from the fit to data. Subsequently, 500 pseudo-experiments are generated based on the same statistical properties as the real data, with the derived PDFs.
 The same fitting method as the nominal result is applied to perform fits on the 500 pseudo-experiments samples.
The difference between the mean value of the fitted signal yields and the nominal signal yield is assigned as the systematic uncertainty of fit bias, amounting to 5.5\%.
To explore the impact of different signal shape models, the signal shape is modified from convolving the MC shape with a Gaussian function to convolving the MC shape with two Gaussian  functions in the alternative fit model. Both the alternative and nominal fits are then applied to the 500 fake data.
The distribution of the differences between the fitted signal yields with these two models describes the deviation between the two fit models. The systematic uncertainty attributed to the signal shape is quantified as the mean value of this distribution, which is 0.7\%.
Similarly, the background shape is modified from the second-order to first-order or third-order Chebyshev polynomial function as an alternative fit model. The systematic uncertainty attributed to the background shape is determined to be 8.8\% using the same method, which is the maximum value of  the two alternative fit model.

\hspace{0.7cm}Similar to Refs.~\cite{BESIII:2023ldd,BESIII:2024wvw}, an event-by-event weighting method is used to study the systematic uncertainty related to the MC generator. 
The signal MC events are weighted according to the momentum distributions of $K^0_S$ and $\Xi^-$ in data.
The deviation between the nominal and reweighting detection efficiencies, 1.7\%, is taken as the systematic uncertainty.
  
\hspace{0.7cm}The systematic uncertainty arising from the size of the signal MC sample is 0.2\%, which is the uncertainty of detection efficiency. 
The uncertainty associated with the total number of $\psi(3686)$ events is 0.5\%~\cite{BESIII:2017tvm}. 
The uncertainties arising from the quoted BFs of $\Xi^- \to \Lambda \pi^-$, $\Lambda \to p \pi^-$ and $K^0_S \to \pi^+\pi^-$ are 0.04\%, 0.8\% and 0.1\%~\cite{PDG}, respectively.

\hspace{0.7cm} The systematic uncertainties are summarized in Table~\ref{tab:systematic}. Assuming that all sources are independent, the total systematic uncertainty on the BF of $\psi(3686) \to \Xi^- K^0_S \bar{\Omega}^+$ is determined to be 11.3\% by adding them in quadrature.

\hspace{0.7cm}The $\bar{\Omega}^+$ signal statistical significance is conservatively estimated to be $5.9\sigma$ after considering the systematic variations of vetoing $\psi(3686) \to \pi^+\pi^- J/\psi$, and the signal and background shapes in the fit to $RM_{\Xi^- K^0_S}$. With considering the systematic effects, the BF of $\psi(3686) \to \Xi^- K^{0}_{S} \bar{\Omega}^+ $ is determined to be $(2.91\pm0.47\pm0.33)\times 10^{-6}$, where the first uncertainty is statistical and the second is systematic.

\begin{table}[htbp]
    \caption{Relative systematic uncertainties in the BF measurement. }
    \label{tab:systematic}
    \begin{center}
    \begin{tabular} {l c c c}
        \hline \hline
        Source & Uncertainty $(\%)$ \\
        \hline
       
       Pion tracking            & 1.0 \\
       Pion PID            & 1.0\\
       $\Lambda$ reconstruction & 2.7 \\
       Mass window and decay length of $\Xi^-$     & 0.2    \\
       Veto $\psi(3686) \to \pi^+\pi^- J/\psi$       & 2.4   \\
       $K^0_S$ reconstruction              &1.2\\
       Signal shape     & 0.7    \\
       Background shape & 8.8   \\ 
       Fit bias & 5.5   \\ 
       MC generator & 1.7       \\
       MC sample size     & 0.2    \\
       $\mathcal{B}_{\Xi^- \to \Lambda \pi^-}$    &   0.1\\
       $\mathcal{B}_{K^0_S \to \pi^+\pi^-}$ & 0.1  \\
       $\mathcal{B}_{\Lambda \to p \pi^-}$           & 0.8    \\
       Total number of $\psi(3686)$ events  & 0.4    \\
        \hline
        Total         & 11.3  \\
        \hline \hline
    \end{tabular}
    \end{center}
\end{table}

\section{Summary}
\label{sec:summary}
In summary, using the world’s largest $\psi(3686)$ data sample taken with the BESIII detector, we observe the $\psi(3686) \to \Xi^- K^0_S \bar{\Omega}^+$ decay for the first time by employing a partial reconstruction method. The measured BF is $(2.91\pm0.47\pm0.33)\times 10^{-6}$, where the first uncertainty is statistical and the second is systematic. This result provides valuable information for understanding the dynamics of $\psi(3686)\to BB^\prime P$ decays. Combining the BF of $\psi(3686) \to \Xi^- K^0_S \bar{\Omega}^+$ measured in this paper and the BF of $\psi(3686) \to \Omega^- K^+ \bar{\Xi}^0$~\cite{BESIII:2024zav}, the ratio $\mathcal{R}=\frac{\mathcal{B}_{\psi(3686) \to \Xi^- K^0_S \bar{\Omega}^+ }}{\mathcal{B}_{\psi(3686) \to \Omega^- K^+ \bar{\Xi}^0}}$ is determined to be $1.05\pm0.23(\rm stat)\pm0.14(\rm syst)$, which deviates from the isospin symmetry conservation predicted value of 0.5 by $2.1\sigma$. It is hard to make any reliable conclusion about this deviation under the current statistics. More precise measurements of these two decays are desirable.

\textbf{Acknowledgement}

The BESIII Collaboration thanks the staff of BEPCII and the IHEP computing center for their strong support. This work is supported in part by National Key R\&D Program of China under Contracts Nos. 2020YFA0406300, 2020YFA0406400, 2023YFA1606000; National Natural Science Foundation of China (NSFC) under Contracts Nos. 11635010, 11735014, 11935015, 11935016, 11935018, 12025502, 12035009, 12035013, 12061131003, 12165022, 12192260, 12192261, 12192262, 12192263, 12192264, 12192265, 12221005, 12225509, 12235017, 12342502, 12361141819; the China Postdoctoral Science Foundation under Grant Number 2024M753040;  the Postdoctoral Fellowship Program of China Postdoctoral Science Foundation under Grant Number GZC20241608; the Chinese Academy of Sciences (CAS) Large-Scale Scientific Facility Program; the CAS Center for Excellence in Particle Physics (CCEPP); Joint Large-Scale Scientific Facility Funds of the NSFC and CAS under Contract No. U1832207; CAS under Contract No. YSBR-101; 100 Talents Program of CAS; the Beijing Natural Science Foundation under Grant Number IS23014; The Institute of Nuclear and Particle Physics (INPAC) and Shanghai Key Laboratory for Particle Physics and Cosmology; Yunnan Fundamental Research Project under Contract No. 202301AT070162; Agencia Nacional de Investigación y Desarrollo de Chile (ANID), Chile under Contract No. ANID PIA/APOYO AFB230003; German Research Foundation DFG under Contract No. FOR5327; Istituto Nazionale di Fisica Nucleare, Italy; Knut and Alice Wallenberg Foundation under Contracts Nos. 2021.0174, 2021.0299; Ministry of Development of Turkey under Contract No. DPT2006K-120470; National Research Foundation of Korea under Contract No. NRF-2022R1A2C1092335; National Science and Technology fund of Mongolia; National Science Research and Innovation Fund (NSRF) via the Program Management Unit for Human Resources \& Institutional Development, Research and Innovation of Thailand under Contract No. B50G670107; Polish National Science Centre under Contract No. 2019/35/O/ST2/02907; Swedish Research Council under Contract No. 2019.04595; The Swedish Foundation for International Cooperation in Research and Higher Education under Contract No. CH2018-7756; U. S. Department of Energy under Contract No. DE-FG02-05ER41374


\bibliographystyle{JHEP}

\providecommand{\href}[2]{#2}\begingroup\raggedright\endgroup

\end{document}